\documentclass[12pt]{article}
\usepackage[numbers]{natbib}
\usepackage{a4}
\usepackage{amsmath}
\usepackage{amssymb}
\usepackage{latexsym}
\usepackage{amssymb}
\usepackage{epsfig}
\usepackage{natbib}
\usepackage{bm}

\usepackage{color}
\def \ii{{\mathrm{i}}}

\def \d{{\mathrm{d}}}

\def \pd{\partial}

\def \e{{\mathrm{e}}}

\def \BJ{{\boldsymbol{J}}}

\def \BB{{\boldsymbol{B}}}
\def \BE{{\boldsymbol{E}}}

\def\onedot{$\mathsurround0pt\ldotp$}
\def\cdddot#1{
  \mathbin{\vcenter{\baselineskip.67ex
    \hbox{\onedot}\hbox{\onedot}\hbox{\onedot}%
  }}%
}

\begin{document}
\title{{\bf
Second gradient electrodynamics:\\
a non-singular relativistic field theory}}
\author{
Markus Lazar$^\text{}$\footnote{
Corresponding author.
{\it E-mail address:} lazar@fkp.tu-darmstadt.de
}
\ and Jakob Leck
\\ \\
        Department of Physics,\\
        Darmstadt University of Technology,\\
        Hochschulstr. 6,\\
        D-64289 Darmstadt, Germany\\
}

\date{\today}
\maketitle

\begin{abstract}
In this paper, we give the covariant formulation of second gradient electrodynamics,
which is a generalized electrodynamics of second order including derivatives of higher order.
The relativistic form of the field equations, the energy-momentum tensor and the Lorentz force density
are presented.
For an electric point charge, the generalized Li\'enard-Wiechert potentials and the corresponding
electromagnetic field strength tensor are given as retarded integral expressions.
Explicit  formulas for the electromagnetic potential vector  and electromagnetic field strength tensor
of a uniformly moving point charge are found without any singularity and discontinuity.
In addition, a world-line integral expression for the self-force of a charged point particle is given.
The relativistic equation of motion of a charged particle coupled with electromagnetic fields in second gradient
electrodynamics is derived, which is an integro-differential equation with nonlocality in time.
For a uniformly accelerated charge,  explicit  formulas of the self-force and the electromagnetic mass, being non-singular, are given.
Moreover, the wave propagation and the dispersion relations in the vacuum of second gradient electrodynamics are analyzed.
Three modes of waves were found:
one non-dispersive wave as in Maxwell electrodynamics, and two dispersive waves similar to the wave propagation in
a collisionless plasma.
\\

\end{abstract}

\newpage

\section{Introduction}

In recent years, 
electromagnetic field theories with higher-order derivatives have been intensively investigated in the literature
(e.g. \citep{Zayats,Perlick2015,Bonin2019,Bonin2019b,Ji,Borges,Hoang,Lazar19}).
The most common theory of this kind is the so-called Bopp-Podolsky theory~\citep{Bopp,Podolsky,PS}
which is the simplest Lorentz and gauge invariant linear generalization of classical electrodynamics. The motivation for the introduction of the theory
was to obtain a consistent theory describing point charges without classical divergences.
Bopp-Podolsky electrodynamics is a linear first-order gradient theory of electrodynamics,
including one length scale parameter, $\ell$, the so-called Bopp-Podolsky parameter, and with fourth-order field equations for the electromagnetic potentials.
The Bopp-Podolsky length scale parameter $\ell$
is of the order of $\sim 10^{-15}$\,m or smaller, therefore of the order of the classical electron radius~(see \citep{Bopp,Iwan,Kvasnica,Cuzi}).
From the mathematical point of view, the length scale parameter $\ell$
plays the role of a regularization parameter.
One of the most remarkable features of the Bopp-Podolsky theory is the fact that it leads to a finite self-energy for a point charge
(see~\citep{Bopp,Podolsky,Lande}).
Moreover, in Bopp-Podolsky electrostatics~\citep{Bopp,Lande3}, the electric potential of a point charge is finite and non-singular,
but the electric field strength is finite and possesses a directional discontinuity at the position of the point charge
which is somewhat unphysical.
In Bopp-Podolsky electrodynamics, the retarded Green function is regular, in fact, it is finite on the light cone,
while its derivative of first order contains a Dirac-delta singularity on the light cone~\citep{Lazar19}.
Eventually, the electromagnetic fields of a non-uniformly moving point charge possess a directional discontinuity on the light cone
(see, e.g.,  \citep{Perlick2015,Ji,Lazar19}), which necessitates a suitable averaging procedure for the electromagnetic field strength tensor
to define the self-field and self-force~\citep{Perlick2015}.
Of course, in a non-singular field theory it would be preferable to obtain non-singular fields and non-singular self-forces without additional definitions or assumptions.

The fact that the first-order derivative of the retarded Green function of the Bopp-Podolsky electrodynamics is still singular on the light cone
was one main motivation to propose second gradient electrodynamics, a linear second-order gradient theory of electrodynamics~\citep{Lazar20}.
In second gradient electrodynamics, the retarded Green function and its derivative of first order are regular.
Second gradient electrodynamics is a linear generalization of
classical Maxwell electrodynamics whose Lagrangian, containing up to   
second-order derivatives of the electromagnetic field strengths, 
is both Lorentz and $U(1)$-gauge invariant.
Using second gradient electromagnetism~\citep{LL20}, the static version of second gradient electrodynamics, it was shown that it
yields non-singular fields for an electric point charge, an electrostatic dipole and a magnetostatic dipole,
and leads to a straightforward regularization of the
dipole-singularities based on higher-order partial differential equations.
In particular, the self-energy of an electrostatic dipole and a magnetostatic dipole is finite, unlike the infinite expressions in
the Bopp-Podolsky electrodynamics~\citep{LL20}.
In \citep{Lazar20}, second gradient electrodynamics was given in the usual vector notation of electromagnetic fields.
In the present work,
the Lagrangian density of second gradient electrodynamics is given in the covariant form by writing it in
the relativistic tensor notation.
From this form, applications to general relativity using second gradient electrodynamics in curved spacetime can be constructed
as it is done in the case of
Bopp-Podolsky electrodynamics
in curved spacetime (see \citep{Zayats16,Cuz18}) or in Bopp-Podolsky cosmology~\citep{Cuzinatto17}.
A relativistic calculation of the self-force is given.
The Lorentz force corresponding to the self-fields of a charged point particle is well-defined and regular in second gradient electrodynamics.
In addition, the covariant form of second gradient electrodynamics can be used as a higher order theory for generalized quantum electrodynamics
as done in the case of Bopp-Podolsky electrodynamics (see, e.g., \citep{Bufalo11,Bufalo12,Bufalo14}).
Moreover, second gradient electrodynamics provides a covariant regularization (spatial and temporal regularization) based on covariant
higher order partial differential equations.
Therefore, covariant second gradient electrodynamics is a suitable candidate for a non-singular relativistic field theory of the electromagnetic fields
of a classical charged point particle. 

On the other hand, the problem of the electromagnetic self-force has a long history going back to the work of Lorentz and Abraham 
modelling charged particles as rigid spheres in the framework of classical electrodynamics.
However, the Abraham-Lorentz model has some important drawbacks: the so-called ``4/3 problem", 
the instability of the charge distribution, and the lack of relativistic covariance.
Dirac~\citep{Dirac} abandoned the rigid sphere model and proposed a model for point-like charged particles 
and he obtained the so-called Lorentz-Dirac equation of motion which is of third order in the time derivative of the position of the particle
and predicts absurd behaviour such as unphysical runaway solutions and pre-acceleration. Moreover, the self-force of a charged point particle 
is singular and not well-defined using classical Maxwell electrodynamics (see, e.g, \citep{Barut,Rohrlich,Spohn,Freeman}). 
In the framework of Bopp-Podolsky electrodynamics, 
the relativistic Lorentz force corresponding to the self-fields of a charged point particle is still not defined in a straightforward way.
In fact, the self-force is defined through averaging, but the result depends on how the averaging is done (see \citep{Perlick2015}).
The solution for the relativistic self-force problem is suggested in the present work using the covariant form of second gradient electrodynamics.
The relativistic Lorentz force and the self-force of a charged point particle are non-singular and well-defined without any need for averaging.
The corresponding equation of motion for a relativistic charged particle moving in its own electromagnetic field is given.
Therefore, the present paper solves the problem of obtaining equations of motion and of the self-force for relativistic charged particles
which are well-defined and singularity-free. 
Such a mathematically consistent modelling of classical charged point particles is of high relevance in accelerator physics  
to describe particle beams in terms of classical particles.
From that point of view, second gradient electrodynamics seems to be a consistent and conceptually well-founded theory of classical charged point particles with non-singular self-force and non-singular self-energy. 

This paper is organized as follows.
In Section~\ref{sec2}, we present the framework of covariant second gradient electrodynamics.
In Section~\ref{sec3}, we derive the energy-momentum tensor and the Lorentz force density in this framework.
The corresponding retarded Green function and its first-order derivative are given in Section~\ref{sec4}.
In Section~\ref{sec5}, we give the generalized Li{\'e}nard-Wiechert potential and electromagnetic field strength tensor
in generalized Li{\'e}nard-Wiechert form.
The self-force and the equation of motion of a charged point particle are given in Section~\ref{sec6}.
The dispersion relations of electromagnetic waves in  second gradient electrodynamics
are presented in Section~\ref{Dispersion}.
Finally, the paper ends with the conclusion in Section~\ref{concl}.
Second gradient electrodynamics in vector form is derived from the covariant form in Appendix~\ref{appendixA}.

\section{Second gradient electrodynamics}
\label{sec2}
\subsection{Notation}

We consider a (four-dimensional) Minkowski space with the metric tensor
\begin{align}
\label{g}
g_{\mu\nu}=g^{\mu\nu}=\begin{pmatrix}
1 & 0 &0 &0\\
0&-1&0&0\\
0&0&-1&0\\
0&0&0&-1
\end{pmatrix}\,.
\end{align}
Here $\mu, \nu,\ldots=0,1,2,3$, and Einstein's summation convention will be used throughout this paper.
In the Minkowski space, the  contravariant components of a four-dimensional space-vector $x$ read
$x^\mu=(ct,\bm x)$,
and the covariant components of the four-vector $x$ read
$x_\mu=(ct,-\bm x)$.
By means of the metric tensor~\eqref{g}, the square of the vector $x$ can be calculated as:
$x^2=g_{\mu\nu}x^\mu x^\nu=c^2 t^2-\bm x^2$.

The speed of light in vacuum is defined by
\begin{align}
c=\frac{1}{\sqrt{\varepsilon_0\mu_0}}\,,
\end{align}
where
$\varepsilon_0$ is the electric constant (or permittivity of vacuum)
and
$\mu_0$ is the magnetic constant
(or permeability of vacuum).

\subsection{Covariant form of second gradient electrodynamics}
\label{Cov}

The covariant formulation of second gradient electrodynamics gives the theory of second gradient electrodynamics
in a form that is manifestly invariant under Lorentz transformations in the formalism of special relativity
using rectilinear inertial coordinate systems.

In second gradient electrodynamics~\citep{Lazar20},
the Lagrangian density reads in covariant form as
\begin{align}
\label{L-BP-rel}
{\cal L}_{\text{grad}}&=
-\frac{1}{4\mu_0}\,
\big(F^{\mu\nu} F_{\mu\nu}
-\ell_1^2\, \pd^{\alpha}F^{\mu\nu} \pd_{\alpha} F_{\mu\nu}
+\ell_2^4\, \pd^{\beta}\pd^{\alpha}F^{\mu\nu} \pd_{\beta}\pd_{\alpha} F_{\mu\nu}
\big)
-j^\nu A_\nu\,.
\end{align}
Here  $\pd_\alpha\equiv\pd/\pd x^\alpha$ denotes the partial derivative,
$A_\nu$ is the electromagnetic potential four-vector (covariant vector), and
\begin{align}
\label{F}
F_{\mu\nu}=\pd_\mu A_\nu-\pd_\nu A_\mu
\end{align}
is the electromagnetic field strength tensor
($F_{\mu\nu}=-F_{\nu\mu}$),
which is a covariant skew-symmetric tensor of rank two.
The contravariant electromagnetic field strength tensor is obtained by raising the indices using the metric tensor~\eqref{g},
\begin{align}
\label{F-trans}
F^{\mu\nu}=g^{\mu\alpha}g^{\nu\beta}F_{\alpha\beta}\,.
\end{align}
$j^\nu$ is the electric current density four-vector.
Moreover,
$\ell_1$ and $\ell_2$ are
the two (positive and real) characteristic length scale parameters in second gradient
electrodynamics, $\ell_1$ is the
length scale parameter corresponding to the first-order derivative and
$\ell_2$ is the length scale parameter corresponding to the second-order derivative
in Eq.~\eqref{L-BP-rel}.

Of course, the electromagnetic field strength tensor~\eqref{F} fulfills the Bianchi identity
\begin{align}
\label{BI-rel}
\pd_\alpha F_{\mu\nu}+\pd_\mu F_{\nu\alpha}+\pd_\nu F_{\alpha\mu}=0\,,
\end{align}
known as the homogeneous Maxwell equations.

The Euler-Lagrange equations derived from the Lagrangian density~\eqref{L-BP-rel}
are given by
\begin{align}
\label{EL-rel}
\frac{\delta {\cal L}_{\text{grad}}}{\delta A_\nu}\equiv
\frac{\pd {\cal L}_{\text{grad}}}{\pd A_\nu}
-\pd_\mu\, \frac{\pd {\cal L}_{\text{grad}}}{\pd(\pd_\mu A_\nu)}
+\pd_\alpha\pd_\mu\, \frac{\pd {\cal L}_{\text{grad}}}{\pd(\pd_\alpha\pd_\mu A_\nu)}
-\pd_\beta\pd_\alpha\pd_\mu\, \frac{\pd {\cal L}_{\text{grad}}}{\pd(\pd_\beta\pd_\alpha\pd_\mu A_\nu)}=0
\end{align}
and represent the inhomogeneous field equations of second gradient electrodynamics,
\begin{align}
\label{EL-rel-F}
 L(\square) \,\pd_\mu F^{\mu\nu}=\mu_0\, j^\nu\,.
\end{align}
Here
the differential operator $L(\square)$ of fourth order is given by
\begin{align}
\label{L-op}
 L(\square)=1+\ell_1^2 \square +\ell_2^4 \square^2\,,
\end{align}
and the d'Alembert operator (or wave operator) is defined as
\begin{align}
\square:=\pd^\mu\pd_\mu\,.
\end{align}
Due to the skewsymmetric property of $F^{\mu\nu}$, the
continuity equation
\begin{align}
\label{CI-rel}
\pd_\nu j^\nu=0\,
\end{align}
follows directly from the field equation~\eqref{EL-rel-F}.

The contravariant electromagnetic excitation tensor is defined as
\begin{align}
\label{CE-rel}
{\cal H}^{\mu\nu}&:=-2\,\frac{\delta{\cal L_{\text{grad}}}}{\delta F_{\mu\nu}}
=\frac{1}{\mu_0}\, L(\square)\, F^{\mu\nu}\,,
\end{align}
which is a skewsymmetric tensor  of rank two, ${\cal H}^{\mu\nu}=-{\cal H}^{\nu\mu}$.
Eq.~\eqref{CE-rel} gives the spacetime relation (or constitutive relation for vacuum) valid in second gradient electrodynamics.
The constitutive law~\eqref{CE-rel} relates the excitation to the field strength,
and it involves a constitutive parameter $\mu_0$ and a constitutive operator $L(\square)$ given in Eq.~\eqref{L-op}.\footnote{In general, constitutive relations involve a set
of constitutive parameters and a set of constitutive operators.
The constitutive operators can be linear and integro-differential in nature, or can imply nonlinear operation on the field (see, e.g., \citep{RC}).}
Therefore, in second gradient electrodynamics the differential operator~\eqref{L-op} enters the constitutive law~\eqref{CE-rel},
giving a weakly nonlocal generalization of the classical constitutive law in Maxwell electrodynamics.
Because the constitutive operator~\eqref{L-op} involves time derivatives and space derivatives, in gradient electrodynamics spacetime is said to be
temporally dispersive and spatially dispersive, respectively.
Temporal dispersion may represent memory effects.
Spatial dispersion on the other hand may represent spreading effects and is significant at small length scales.
Using the electromagnetic excitation tensor~\eqref{CE-rel}, the inhomogeneous field equation~\eqref{EL-rel-F} reduces to
a ``Maxwell" form
\begin{align}
\label{EL-rel-H}
\pd_\mu {\cal H}^{\mu\nu}= j^\nu\,.
\end{align}

If we use the decomposition of the
electromagnetic excitation tensor (see, e.g., \citep{Post})
\begin{align}
\label{CE-rel-2}
{\cal H}^{\mu\nu}&=\frac{1}{\mu_0}\, F^{\mu\nu}+{\cal M}^{\mu\nu}\,
\end{align}
into the classical vacuum field $\frac{1}{\mu_0}\, F^{\mu\nu}$ and a tensor ${\cal M}^{\mu\nu}$ of electric and magnetic polarization, we obtain for the tensor of
electric and magnetic polarization present in the vacuum of second gradient electrodynamics
\begin{align}
\label{M}
{\cal M}^{\mu\nu}=\frac{1}{\mu_0}\,
\big[\ell_1^2 \square +\ell_2^4 \square^2\big] F^{\mu\nu}\,.
\end{align}
In other words, the differential operator $\big[\ell_1^2 \square +\ell_2^4 \square^2\big]$ gives rise to
electric and magnetic polarization of the vacuum.

Using the generalized Lorenz gauge condition~\citep{GP,Lazar20} \footnote{The generalized Lorenz gauge is necessary for a quantization of the theory. Here the classical Lorenz gauge would yield the same results.}
\begin{align}
\label{EL-rel-LG}
L(\square) \, \pd_\mu A^{\mu}=0\,,
\end{align}
the electromagnetic potential satisfies an inhomogeneous partial differential equation of sixth order,
\begin{align}
\label{EL-rel-A}
L(\square)\, \square A^{\nu}=\mu_0\, j^\nu\,.
\end{align}
Using Eqs.~\eqref{F} and \eqref{EL-rel-A},
it can be seen that the electromagnetic field strength tensor fulfills the inhomogeneous sixth-order partial differential equation
\begin{align}
\label{EL-rel-F-2}
L(\square)\, \square F^{\mu\nu}=\mu_0  \big(\pd^\mu j^\nu-\pd^\nu j^\mu\big)\,.
\end{align}

The differential operator of fourth order~\eqref{L-op}
can be written as a product of two Klein-Gordon operators with
two length scale parameters $a_1$ and $a_2$,
which is called bi-Klein-Gordon operator,
\begin{align}
\label{L-op-2}
L(\square)=\big(1+a_1^2\square\big)\big(1+a_2^2\square\big)
\end{align}
with
\begin{align}
\label{a1a2-1}
\ell_1^{2}&=a_1^{2}+a_2^{2}\, ,\\
\label{a1a2-2}
\ell_2^{4}&=a_1^{2}\, a_2^{2}\,
\end{align}
and
\begin{align}
\label{a1-2}
a^{2}_{1,2}&=\frac{\ell_1^{2}}{2}\Bigg(1\pm\sqrt{1-4\,\frac{\ell_2^{4}}{\ell_1^{4}}}\Bigg)\,.
\end{align}
If $\ell_1^4>4\ell_2^4$, then
$a_1$ and $a_2$ are real and distinct and they read as
\begin{align}
\label{a1-2-2}
a_{1,2}&=\ell_1\,\sqrt{\frac{1}{2}\pm \frac{1}{2}\,\sqrt{1-4\left(\frac{\ell_2}{\ell_1}\right)^{\!4}}}\,,
\end{align}
we choose $a_1>a_2$.
The limit to Bopp-Podolsky electrodynamics  is
$\ell_2^4\rightarrow 0$, and therefore $a_1\rightarrow \ell$ and
$a_2\rightarrow 0$.

\section{Energy-momentum tensor and Lorentz force density}
\label{sec3}

In the classical Maxwell electrodynamics (see, e.g., \citep{Jackson,Post,Barut,Rohrlich})
as well as in the Bopp-Podolsky theory (see, e.g., \citep{Acc,Perlick2015,Hoang}),
the Lorentz force density is given by the divergence of the energy-momentum tensor.
Therefore, the energy-momentum tensor of second gradient electrodynamics is to be determined.

The procedure of deriving the Lorentz force density is based on the Noether theorem and broken translation symmetry.
The translation symmetry  will be broken by the current density $j_\mu$ appearing in the Lorentz force density.
In order to calculate the energy-momentum tensor and the Lorentz force density,
we start with the free-field Lagrangian density of second gradient electrodynamics
\begin{align}
\label{L-F}
{\cal L}_F&=
-\frac{1}{4\mu_0}\,
\big(F^{\mu\nu} F_{\mu\nu}
-\ell_1^2\, \pd^{\alpha}F^{\mu\nu} \pd_{\alpha} F_{\mu\nu}
+\ell_2^4\, \pd^{\beta}\pd^{\alpha}F^{\mu\nu} \pd_{\beta}\pd_{\alpha} F_{\mu\nu}
\big)\,.
\end{align}
Because the  Lorentz force density is related to the four-dimensional translation in spacetime,
we consider the functional derivative to be translational
\begin{align}
\label{delta-tr}
\delta=\varepsilon^\lambda \pd_\lambda\,,
\end{align}
where $\pd_\lambda$ is the infinitesimal generator of translation in spacetime and $\varepsilon^\lambda$ is a constant translation in $x^\lambda$-direction.
The infinitesimal variation of the free-field Lagrangian~\eqref{L-F} reads as
\begin{align}
\delta {\cal L}_F
=\varepsilon^\lambda \pd_\lambda
{\cal L}_F\,.
\end{align}

On the one hand, the action of the infinitesimal generator of translation on the free field Lagrangian density becomes
\begin{align}
\label{N-1}
\pd_\lambda {\cal L}_F&=
-\frac{1}{2\mu_0}\,
\big(F^{\mu\nu} \pd_\lambda F_{\mu\nu}
-\ell_1^2\, \pd^{\alpha}F^{\mu\nu} \pd_\lambda  \pd_{\alpha} F_{\mu\nu}
+\ell_2^4\, \pd^{\beta}\pd^{\alpha}F^{\mu\nu}  \pd_\lambda \pd_{\beta}\pd_{\alpha} F_{\mu\nu}
\big)
\nonumber\\
&=
-\frac{1}{\mu_0}\,
\big(F^{\mu\nu} \pd_\nu F_{\mu\lambda}
-\ell_1^2\, \pd^{\alpha}F^{\mu\nu} \pd_\nu  \pd_{\alpha} F_{\mu\lambda}
+\ell_2^4\, \pd^{\beta}\pd^{\alpha}F^{\mu\nu}  \pd_\nu \pd_{\beta}\pd_{\alpha} F_{\mu\lambda}
\big)\,,
\end{align}
where the Bianchi identity~\eqref{BI-rel} has been used.

There are the following identities
\begin{align}
\label{N-rel1}
F^{\mu\nu} \pd_\nu F_{\mu\lambda}=-\pd_\nu F^{\mu\nu}F_{\mu\lambda}
+\pd_\nu\big[F^{\mu\nu}F_{\mu\lambda}\big]\,,
\end{align}
\begin{align}
\label{N-rel2}
\pd^\alpha F^{\mu\nu} \pd_\nu \pd_\alpha F_{\mu\lambda}=\square \pd_\nu F^{\mu\nu}F_{\mu\lambda}
+\pd_\nu\big[\pd^\nu F^{\mu\alpha} \pd_\alpha F_{\mu\lambda} -\square F^{\mu\nu}F_{\mu\lambda}\big]
\end{align}
and
\begin{align}
\label{N-rel3}
\pd^\beta\pd^\alpha F^{\mu\nu} \pd_\nu \pd_\beta \pd_\alpha F_{\mu\lambda}=
-\square^2 \pd_\nu F^{\mu\nu}F_{\mu\lambda}
+\pd_\nu\big[\pd^\nu \pd^\beta F^{\mu\alpha} \pd_\alpha \pd_\beta F_{\mu\lambda} -\square  \pd^\nu F^{\mu\alpha}\pd_\alpha F_{\mu\lambda}
+\square^2 F^{\mu\nu}F_{\mu\lambda}
\big]\,.
\end{align}

On the other hand, the action of the translational generator on the free field Lagrangian density can be written as
\begin{align}
\label{N-2}
\pd_\lambda {\cal L}_F=\pd_\nu  \delta^\nu_{\lambda}{\cal L}_F\,,
\end{align}
where $g^\nu_{\ \, \lambda}\equiv\delta^\nu_{\lambda}$.

By substituting Eqs.~\eqref{N-rel1}--\eqref{N-2} into Eq.~\eqref{N-1}, we obtain
\begin{align}
\label{N-3}
&\pd_\nu \bigg(
-{\cal L}_F \, \delta^\nu_{\lambda}
-\frac{1}{\mu_0}\,
L(\square) F^{\mu\nu}F_{\mu\lambda}
+\frac{1}{\mu_0}\, \big[\ell_1^2+\ell_2^4\square\big] \pd^\nu F^{\mu\alpha} \pd_\alpha F_{\mu\lambda}
-\frac{1}{\mu_0}\, \ell_2^4\,  \pd^\nu \pd^\beta F^{\mu\alpha} \pd_\alpha \pd_\beta F_{\mu\lambda}\bigg)
\nonumber\\
&\qquad
=-\frac{1}{\mu_0}\,
L(\square)\,  \pd_\nu F^{\mu\nu}F_{\mu\lambda}\,.
\end{align}

In the divergence-term of Eq.~\eqref{N-3},
the energy-momentum tensor of the electromagnetic field in the theory of second gradient electrodynamics can be defined as
\begin{align}
\label{EMT}
T^\nu_{\ \, \lambda}
=-{\cal L}_F\,  \delta^\nu_{\lambda}
-\frac{1}{\mu_0}\,
L(\square) F^{\mu\nu}F_{\mu\lambda}
+\frac{1}{\mu_0}\, \big[\ell_1^2+\ell_2^4\square\big] \pd^\nu F^{\mu\alpha} \pd_\alpha F_{\mu\lambda}
-\frac{1}{\mu_0}\, \ell_2^4 \, \pd^\nu \pd^\beta F^{\mu\alpha} \pd_\alpha \pd_\beta F_{\mu\lambda}\,,
\end{align}
which is gauge invariant and an asymmetric tensor.
Using Eq.~\eqref{CE-rel}, the energy-momentum tensor~\eqref{EMT} reduces to
\begin{align}
\label{EMT-2}
T^\nu_{\ \, \lambda}
=-{\cal L}_F\,  \delta^\nu_{\lambda}
-\mathcal{H}^{\mu\nu}F_{\mu\lambda}
+\frac{1}{\mu_0}\, \big[\ell_1^2+\ell_2^4\square\big] \pd^\nu F^{\mu\alpha} \pd_\alpha F_{\mu\lambda}
-\frac{1}{\mu_0}\, \ell_2^4 \, \pd^\nu \pd^\beta F^{\mu\alpha} \pd_\alpha \pd_\beta F_{\mu\lambda}\,.
\end{align}
So, the divergence of the energy-momentum tensor gives the following translational balance law
\begin{align}
\label{EMT-3}
\pd_\nu T^\nu_{\ \, \lambda}=j^\mu F_{\mu\lambda}\,,
\end{align}
where Eq.~\eqref{EL-rel-F} has been used.
The contravariant energy-momentum tensor can be obtained by using
\begin{align}
\label{EMT-4}
T^{\nu\kappa}=g^{\kappa\lambda} T^\nu_{\ \, \lambda}
\end{align}
and the divergence of the energy-momentum tensor gives the Lorentz force density
\begin{align}
\label{EMT-5}
\pd_\nu T^{\nu\kappa}=-F^{\kappa\mu}j_\mu\,.
\end{align}
Finally, the Lorentz force density four-vector reads
\begin{align}
\label{LF-d}
f^\kappa=F^{\kappa\mu}j_\mu\,.
\end{align}
Therefore,
the force density in second gradient electrodynamics is precisely the Lorentz force density~\eqref{LF-d}.

If $j_\mu=0$, then Eq.~\eqref{EMT-2}
gives a conservation law, due to translation symmetry,
 \begin{align}
\label{EMT-cl}
\pd_\nu T^{\nu\kappa}=0\,.
\end{align}

\section{Retarded Green function and solutions of the covariant field equations}
\label{sec4}

\begin{figure}[t]\unitlength1cm
\vspace*{0.1cm}
\centerline{
\epsfig{figure=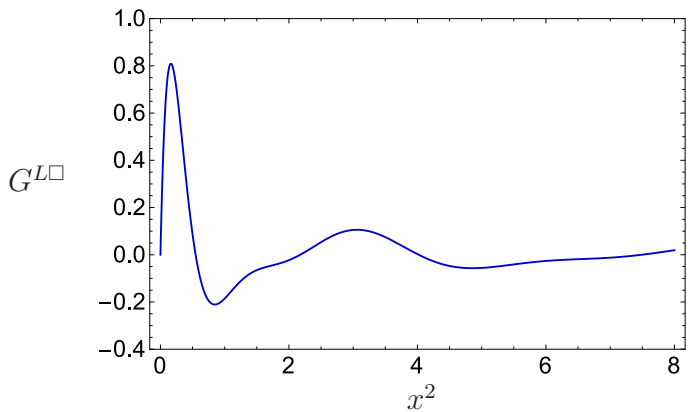,width=7.8cm}
\put(-3.7,-0.4){$x^2 $}
\put(-9,2.5){$G^{L\square} $}
}
\caption{Green function $G^{L\square}$ plotted against $x^2$ ($a_1=0.2$ and $a_2=0.1$).}
\label{fig:GF}
\end{figure}

The Green function (fundamental solution) of the bi-Klein-Gordon-d'Alembert equation,
which is a partial differential equation of sixth order, is defined via
\begin{align}
\label{BPE}
 L(\square)\,
\square\, G^{L\square}(x)=\delta^4(x)\,,
\end{align}
where $\delta^4$ is the four-dimensional Dirac delta-function.
The retarded solution of Eq.~\eqref{BPE} reads as (see also \citep{Lazar20})
\begin{align}
\label{G-LD}
G^{L\square}(x)
&=\frac{1}{4\pi(a_1^2-a_2^2)}\,
\frac{H\big(x^0\big)H\big(x^2\big)}{\sqrt{x^2}}\,
\bigg[
{a}_1 J_1 \bigg(\frac{\sqrt{x^2}}{a_1}\bigg)
-a_2 J_1 \bigg( \frac{\sqrt{x^2}}{a_2}\bigg)
\bigg]\,.
\end{align}
The retarded Green function is the only Green function (fundamental solution) of the hyperbolic differential operator,
$ L(\square) \square$, with support in the half-space  $ t\ge 0$.
Here $H$ is the Heaviside step function
and $J_1$ is the Bessel function of the first kind of order 1.
Taking into account that
\begin{align}
\label{rel-J1}
\lim_{z \to 0} \,\frac{1}{z}\,J_1(z)=\frac{1}{2}\,,
\end{align}
we obtain that the value of
the Green function~\eqref{G-LD} is zero on the light cone $x^2=0$.
The Green function~\eqref{G-LD}
depends on the coordinate $x$ only by means of the invariant $x^2$.
Note that the factor $H\big(x^0\big)$ is not invariant under Lorentz transformations, but the product $H\big(x^0\big)H\big(x^2\big) = H\big(x^0 - |\bm x|\big)$ is invariant under orthochronous Lorentz transformations.
The Green function~\eqref{G-LD} vanishes  outside the light cone, where $x^2<0$.
The retarded Green function~\eqref{G-LD} behaves proportional to $J_1(z)/z$ inside the forward light cone $x^2>0$ ($t\ge 0$).
The retarded Green function~\eqref{G-LD} is plotted against $x^2$ in Fig.~\ref{fig:GF}.
In Fig.~\ref{fig:GF}, it can be seen that  the Green function~\eqref{G-LD} decays rather than oscillates.

\begin{figure}[t]\unitlength1cm
\vspace*{0.1cm}
\centerline{
\epsfig{figure=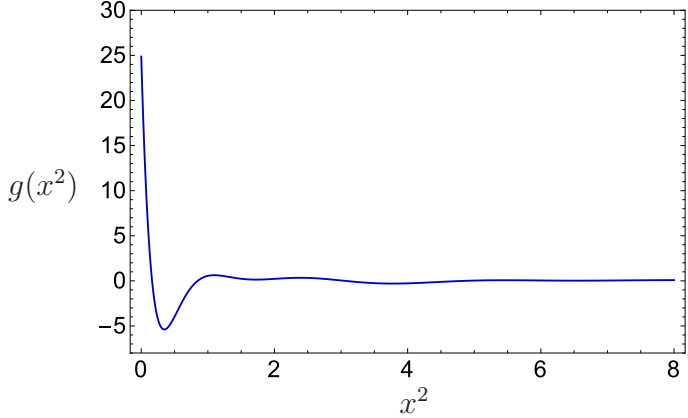,width=7.8cm}
\put(-3.8,-0.4){$x^2 $}
\put(-9,2.5){$g(x^2)$}
}
\caption{$g(x^2)$ plotted against $x^2$ ($a_1=0.2$ and $a_2=0.1$).}
\label{fig:GF-grad}
\end{figure}

For the first-order derivative of the Green function~\eqref{G-LD} we obtain
\begin{align}
\label{G-LD-grad}
\pd^\mu G^{L\square}(x)
&=-\frac{x^\mu}{4\pi(a_1^2-a_2^2)}\,
\frac{H\big(x^0\big)H\big(x^2\big)}{x^2}
\bigg[
J_2 \bigg(\frac{\sqrt{x^2}}{a_1}\bigg)
-J_2 \bigg(\frac{\sqrt{x^2}}{a_2}\bigg)
\bigg]\,.
\end{align}
Here the relation $(J_1(z)/z)'=-J_2(z)/z$ has been used, and
$J_2$ is the Bessel function of the first kind of order 2.
Taking into account that
\begin{align}
\label{J2-rel}
\lim_{z \to 0} \,\frac{1}{z^2}\,J_2(z)=\frac{1}{8}\,,
\end{align}
it can be observed that the first-order derivative of the Green function~$G^{L\square}$
is finite on the light cone, namely
\begin{align}
\pd^\mu G^{L\square}(x)
=
\frac{x^\mu }{32\pi a_1^2a_2^2}\,H\big(x^0\big)H\big(x^2\big)\,\big(1 + \mathcal{O}\big(x^2\big)\big)\,.
\end{align}
Eq.~\eqref{G-LD-grad} can be written in the form
\begin{align}
\label{G-grad-deco}
\pd^\mu G^{L\square}(x)
&=x^\mu g(x^2)\,,
\end{align}
where the part $g(x^2)$ is the part of $\pd^\mu G^{L\square}$
depending on $x^2$ and is plotted against $x^2$ in Fig.~\ref{fig:GF-grad}.
In Fig.~\ref{fig:GF-grad}, it can be seen that $g(x^2)$ is finite on the light cone, $x^2=0$, and it falls rapidly away from the light cone.

Thus, we have the result that the Green function $G^{L\square}(x)$ and its derivative of first order are regular.
On the light cone, the Green function $G^{L\square}(x)$ is zero and its
first-order derivative is finite there.

Using the Green function~\eqref{G-LD}, the
retarded solution of Eq.~\eqref{EL-rel-A} is given by the convolution
\begin{align}
\label{A-rp}
A^\mu &=\mu_0\, G^{L\square}* j^\mu\,.
\end{align}
The symbol $*$ denotes the convolution in spacetime.
So,  the retarded electromagnetic potential four-vector reads as
\begin{align}
\label{A-rp2}
A^\mu(x) &=\mu_0 \int\limits_{\Bbb R^4} G^{L\square}(x-x') \,j^\mu(x')\, \d^4 x'\,,
\end{align}
where $\d^4 x=c\,\d t\, \d V$.
The
retarded solution of Eq.~\eqref{EL-rel-F} is given by the convolution
\begin{align}
\label{F-rp}
F^{\mu\nu} &=\mu_0\, G^{L\square}* \big(\pd^\mu j^\nu-\pd^\nu j^\mu\big)\nonumber\\
&=\mu_0\, \big(\pd^\mu G^{L\square}* j^\nu-\pd^\nu  G^{L\square}* j^\mu\big)\,.
\end{align}
Therefore, the retarded electromagnetic field strength tensor reads as
\begin{align}
\label{F-rp2}
F^{\mu\nu}(x) &=\mu_0\int\limits_{\Bbb R^4}
\big[\pd^\mu G^{L\square}(x-x')\, j^\nu(x')-\pd^\nu  G^{L\square}(x-x')\, j^\mu(x')\big]\,
\d^4 x'
\,.
\end{align}

\section{A non-uniformly moving charged point particle}
\label{sec5}

For a point charge with world line $\xi(\tau)$, the current density is given by
\begin{align}
\label{j-loop}
j^\mu(x)= qc \int\limits_{-\infty}^{\ \infty} u^\mu(\tau) \, \delta^4 (x-\xi(\tau))\, \d \tau\,,
\end{align}
where $\tau$ is the proper time, $u^\mu(\tau) =\d \xi^\mu(\tau)/\d \tau$  is the four-velocity of the particle and $q$ is the electric charge.
Thus, $j^\mu$ is concentrated on the world line of the point charge.

If we substitute Eqs.~\eqref{G-LD} and \eqref{j-loop} into Eq.~\eqref{A-rp2} and carry out the $x'$-integration,
we obtain for the electromagnetic potential four-vector of a non-uniformly moving charged point particle
\begin{align}
\label{A-LW}
A^\mu(x)
&=\frac{\mu_0 qc}{4\pi(a_1^2-a_2^2)}\int\limits_{-\infty}^{\tau_R}
\frac{u^\mu(\tau)}{\sqrt{R^2(\tau)}}\,
\bigg[
a_1 J_1 \bigg(\frac{\sqrt{R^2(\tau)}}{a_1}\bigg)
-a_2 J_1 \bigg(\frac{\sqrt{R^2(\tau)}}{a_2}\bigg)
\bigg]\, \d \tau\,,
\end{align}
where $R^\mu(\tau)=x^\mu-\xi^\mu(\tau)$. The integration is carried out over the range where
\begin{align}
\label{tau-r1}
 x^0 - \xi^0(\tau) > 0\, , \quad (x - \xi(\tau))^2 > 0
\end{align}
or, equivalently,
\begin{align}
\label{tau-r2}
 x^0 - \xi^0(\tau) > |\bm{x} - \bm{\xi}(\tau)| \,,
\end{align}
and the retarded time $\tau_R$ is defined as the value of the proper time for which equality holds in Eq.~\eqref{tau-r2} (see also \citep{Lazar20}), i.e. where the world-line intersects the backwards light cone from $x^\mu$.
Note that Eq.~\eqref{A-LW} represents the generalized Li\'enard-Wiechert potential valid in second gradient electrodynamics.

By substituting Eqs.~\eqref{G-LD-grad} and \eqref{j-loop} into Eq.~\eqref{F-rp2} and performing the $x'$-integration,
the electromagnetic field strength tensor of a non-uniformly moving charged point particle is obtained as\footnote{Using the notation
of  \citet{Schouten}, brackets denote
antisymmmetrization: $A^{[\nu}B^{\mu]}=(A^\nu B^\mu-A^\mu B^\nu)/2!$.}
\begin{align}
\label{F-LW}
F^{\mu\nu}(x)
&=\frac{\mu_0 qc}{2\pi (a_1^2-a_2^2)}
\int\limits_{-\infty}^{\tau_R}
\frac{R^{[\nu}(\tau)u^{\mu]}(\tau)
}{R^2(\tau)}\,
\bigg[
J_2 \bigg(\frac{\sqrt{R^2(\tau)}}{a_1}\bigg)
-J_2 \bigg(\frac{\sqrt{R^2(\tau)}}{a_2}\bigg)
\bigg]\, \d\tau\,.
\end{align}
Eq.~\eqref{F-LW} represents the electromagnetic field strength of generalized Li\'enard-Wiechert type valid in second gradient electrodynamics.
Both the electromagnetic potential~\eqref{A-LW} and the electromagnetic field strength~\eqref{F-LW}
depend on the entire earlier history of the non-uniformly moving point charge up to the retarded time $\tau_R$.
Therefore, a point charge is haunted by its past.
Note that the electromagnetic field strength~\eqref{F-LW} does not possess a strong directional dependence
which is present in the Bopp-Podolsky electrodynamics (see \citep{Perlick2015,Lazar19,Hoang}).
\\

\begin{figure}[t]\unitlength1cm
\vspace*{0.1cm}
\centerline{
\epsfig{figure=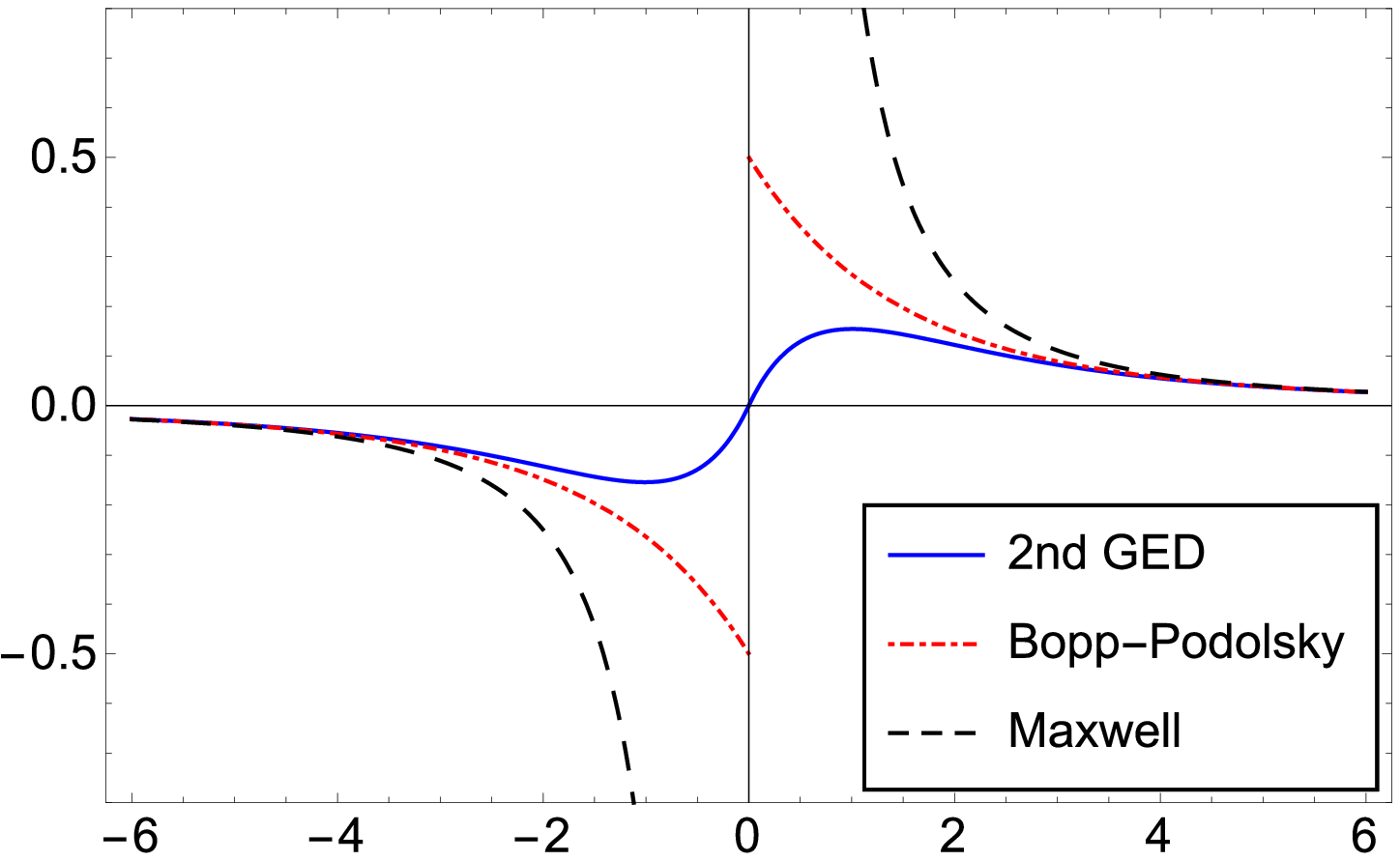,width=8cm}
\put(-4.1,-0.4){$x_1/\ell $}
\put(-8.8,2.5){$\displaystyle{\frac{E_1}{E_0}}$}
}
\caption{$E_1$-component of the electrostatic field strength of a point charge in Maxwell, Bopp-Podolsky  and 
second gradient electrodynamics (2nd GED) along the $x_1$-axis. The electric field is plotted 
in units of $E_0 = \frac{q}{4 \pi \varepsilon_0 \ell^2}$ for $\ell = a_1 = 2 a_2$.}
\label{fig:E-fields}
\end{figure}

\noindent
\textit{Example: Charge at rest}\\
For a charge at rest in a suitably chosen inertial frame the above integrals can be computed explicitly (cf. example 1 in \citep{Perlick2015} in Bopp-Podolsky electrodynamics). The world line reads
\begin{align}
\label{car-wl}
 \xi^\mu(\tau) = V^\mu \tau
\end{align}
with $V_\mu V^\mu = c^2$ and $V^0 > 0$. The retarded time is found to be
\begin{align}
\label{car-rt}
 \tau_R = \frac{1}{c^2} \bigg(x^\mu V_\mu - \sqrt{(x^\mu V_\mu)^2 - x^2 c^2}\bigg)\,,
\end{align}
and reparametrizing with the coordinate
\begin{align}
\label{car-zeta}
 \zeta = \sqrt{R^2(\tau)}\,,
\end{align}
running from $\zeta = 0$ at $\tau_R$ to $\zeta \rightarrow \infty$ for $\tau \rightarrow - \infty$ , integrals are found to simplify as
\begin{align}
\label{car-int}
 \int\limits_{-\infty}^{\tau_R} f\big(R^2(\tau)\big)\,\text{d}\tau = \frac{1}{c} \int\limits_0^\infty f(\zeta^2) \frac{\zeta}{\sqrt{r_R^2(x) + \zeta^2}}\,\text{d}\zeta\,
\end{align}
for some function $f$. Here, $r_R$ denotes the retarded distance with
\begin{align}
 \label{car-rr}
 c r_R(x)= u^\mu(\tau_R) \big(x_\mu - \xi_\mu(\tau_R)\big) = V^\mu x_\mu - c^2 \tau_R\,.
\end{align}
Applying this to the four-potential \eqref{A-LW} yields
\begin{align}
 \label{A-car-int}
 A^\mu(x)
&=\frac{\mu_0 q \,V^\mu}{4\pi(a_1^2-a_2^2)}\int\limits_0^\infty \bigg[
a_1 J_1 \bigg(\frac{\zeta}{a_1}\bigg)
-a_2 J_1 \bigg(\frac{\zeta}{a_2}\bigg)
\bigg] \frac{1}{\sqrt{r_R^2(x)+\zeta^2}}\,\text{d}\zeta
\end{align}
and after evaluation of the integral, using (6.552-1), (8.467) and (8.469-3) in \citep{GR}, we find
\begin{align}
\label{A-car}
 A^\mu(x)
&=\frac{\mu_0 q\, V^\mu}{4\pi r_R(x)}\bigg(1 - \frac{1}{a_1^2-a_2^2}\,
\Big[a_1^2 \e^{-r_R(x)/a_1}-a_2^2 \e^{-r_R(x)/a_2}\Big] \bigg)\,,
\end{align}
which is finite for all $x$.  In the rest system of the charge with $V^\mu = (c,0,0,0)$, $r_R(x) = |\bm x|$ is the Euclidean distance and the result reduces to the electrostatic solution given in \citep{LL20}.
For the field strength tensor we note that in the example of a charge at rest $R^{[\nu}(\tau)u^{\mu]}(\tau) = x^{[\nu}V^{\mu]}$ is independent of the integration variable,
and thus Eq.~\eqref{F-LW} becomes
\begin{align}
 \label{F-car-int}
F^{\mu\nu}(x)
&=\frac{\mu_0 q c\, x^{[\nu}V^{\mu]}}{2\pi (a_1^2-a_2^2)}
\int\limits_{-\infty}^{\tau_R}
\frac{1}{R^2(\tau)}\,
\bigg[
J_2 \bigg(\frac{\sqrt{R^2(\tau)}}{a_1}\bigg)
-J_2 \bigg(\frac{\sqrt{R^2(\tau)}}{a_2}\bigg)
\bigg]\, \d\tau \nonumber\\
&=\frac{\mu_0 q c\, x^{[\nu}V^{\mu]}}{2\pi (a_1^2-a_2^2)} \int\limits_0^\infty \bigg[
J_2 \bigg(\frac{\zeta}{a_1}\bigg)
-J_2 \bigg(\frac{\zeta}{a_2}\bigg)
\bigg] \frac{1}{\zeta \sqrt{r_R^2(x)+\zeta^2}}\,\text{d}\zeta \,.
\end{align}
Evaluating the integral with the help of (8.471-1), (6.552-1), (8.467) and (8.468) in \citep{GR}, we obtain
\begin{align}
 \label{F-car}
 F^{\mu\nu}(x)
=\frac{\mu_0 q \, x^{[\nu}V^{\mu]}}{2\pi r_R^3(x)} \bigg[1
&-\frac{1}{a_1^2-a_2^2}\,
\Big[a_1^2 \e^{-r_R(x)/a_1}-a_2^2 \e^{-r_R(x)/a_2}\Big]\nonumber\\
&-\frac{r_R(x)}{a_1^2-a_2^2}\,
\Big[a_1 \e^{-r_R(x)/a_1}-a_2 \e^{-r_R(x)/a_2}\Big]\bigg]\,,
\end{align}
which is finite everywhere.
While in the Bopp-Podolsky theory the corresponding expression
has a directional discontinuity on the world line,
we here find that the limit where $x$ approaches the world line exists
and is zero. Again, in the rest system of the charge
the result reduces to the electrostatic solution
\begin{align}
\label{E-car}
\bm E &= \frac{q}{4\pi\varepsilon_0}\, \frac{\bm x}{r^3}\,\bigg[1-\frac{1}{a_1^2-a_2^2}\,
\Big[a_1^2 \e^{-r/a_1}-a_2^2 \e^{-r/a_2}\Big]
-\frac{r}{a_1^2-a_2^2}\,
\Big[a_1 \e^{-r/a_1}-a_2 \e^{-r/a_2}\Big]\bigg]\nonumber\\
&= \frac{q}{4\pi\varepsilon_0}\,\frac{\bm x}{3 a_1 a_2 (a_1 + a_2)} + \mathcal{O}\big(r^2\big)
\end{align}
with $r=|\bm x|$ given in \citep{LL20}. The electric field is zero
at the location of the charge, see Fig.~\ref{fig:E-fields}.
Note that the terms appearing in Eqs.~\eqref{A-car} and \eqref{F-car} have
the same structure as and could be written in terms of
the auxiliary functions $f_0$ and $f_1$ in \citep{LL20}.

\section{Self-force and equation of motion}
\label{sec6}

Using the Lorentz force density~\eqref{LF-d} and the current density~\eqref{j-loop},
the Lorentz force four-vector is given by
\begin{align}
\label{F-force}
{\mathcal F}^\mu=q F^{\mu\nu}u_{\nu}\,,
\end{align}
which is the four-force on a moving charge $q$ with four-velocity $u_\nu$ in presence of the  electromagnetic field strength $F^{\mu\nu}$.

If a non-uniformly moving charge $q$ interacts with its own electromagnetic field $F^{\mu\nu}$, we speak of a self-force.
The self-force four-vector of a non-uniformly moving point charge is obtained as
\begin{align}
\label{F-sf}
{\mathcal F}^\mu_{\text{sf}}(\tau)
&=\frac{\mu_0 q^2c\, u_\nu(\tau)}{2\pi (a_1^2-a_2^2)}
\int\limits_{-\infty}^{\ \tau}
\frac{R^{[\nu}(\tau,\tau')u^{\mu]}(\tau')}{R^2(\tau,\tau')}\,
\bigg[
J_2 \bigg(\frac{\sqrt{R^2(\tau,\tau')}}{a_1}\bigg)
-J_2 \bigg(\frac{\sqrt{R^2(\tau,\tau')}}{a_2}\bigg)
\bigg]\, \d\tau'\,,
\end{align}
where $R^\mu(\tau,\tau')=\xi^\mu(\tau)-\xi^\mu(\tau')$.
The self-force of a point charge in second gradient electrodynamics, Eq.~\eqref{F-sf}, is well-defined on the world line,
and no averaging procedures like in Bopp-Podolsky electrodynamics \citep{Perlick2015} are necessary for its definition.

The equation of motion of a charged particle
which is under the influence of an external force ${\mathcal F}^\mu_{\text{ext}}$
taking into account self-force effects reads as
\begin{align}
\label{EOM}
m  a^\mu(\tau)
= {\mathcal F}^\mu_{\text{sf}}(\tau)
+  {\mathcal F}^\mu_{\text{ext}}(\tau)\,,
\end{align}
where $m$ is the ``bare'' mass (or mechanical mass) of the particle,
$a^\mu(\tau) =\d^2 \xi^\mu(\tau)/\d \tau^2$
denotes the four-vector of its acceleration,
and  ${\mathcal F}^\mu_{\text{sf}}$ is the self-force.
Eq.~\eqref{EOM} is an integro-differential equation for $\xi(\tau)$ and is nonlocal in time
due to the self-force~\eqref{F-sf}.
No infamous third-order time derivative of $\xi(\tau)$ shows up in the equation of motion of a charged particle in
second gradient electrodynamics unlike in the Lorentz-Dirac equation in classical Maxwell electrodynamics.
\\

\noindent
\textit{Example: Uniformly accelerated charge}\\
To compare with the results of \citet{Zayats} and \citet{Perlick2015} obtained in Bopp-Podolsky electrodynamics,
we consider a charge with constant acceleration $a$ in its instantaneous rest frame with the world line
\begin{align}
 \label{acc-wl}
 \xi^\mu(\tau) = \frac{c^2}{a} \bigg(\sinh\Big(\frac{a\tau}{c}\Big),\cosh\Big(\frac{a\tau}{c}\Big),0,0\bigg)\,.
\end{align}
To simplify the integral \eqref{F-sf} we again use the coordinate
\begin{align}
 \label{acc-zeta}
 \zeta = \sqrt{R^2(\tau, \tau')}
\end{align}
for which we find
\begin{align}
\label{acc-zeta-square}
 \zeta^2 = \frac{2 c^4}{a^2}\bigg[\cosh\Big(\frac{a}{c}(\tau-\tau')\Big) - 1\bigg]
\end{align}
and
\begin{align}
\label{acc-diff}
 c\,\text{d}\tau' = - \frac{1}{\sqrt{\frac{\zeta^2 a^2}{4 c^4}+1}}\, \text{d}\zeta\,.
\end{align}
With
\begin{align}
 \label{acc-B}
 u_\nu(\tau) R^{[\nu}(\tau,\tau')u^{\mu]}(\tau') = - \frac{1}{4} \zeta^2 a^\mu(\tau)
\end{align}
the self-force integral reduces to
\begin{align}
 \label{acc-sf-int}
 {\mathcal F}^\mu_{\text{sf}}(\tau)
&=-\frac{\mu_0 q^2\, a^\mu(\tau)}{8\pi (a_1^2-a_2^2)}
\int\limits_{0}^{\infty}
\frac{J_2 \big(\frac{\zeta}{a_1}\big)
-J_2 \big(\frac{\zeta}{a_2}\big)
}{\sqrt{\frac{\zeta^2 a^2}{4 c^4}+1}}\, \d\zeta\,,
\end{align}
and, again using (6.552-1) in \citep{GR}, with the modified Bessel functions of first and second kind $I_1$ and $K_1$ the self-force can be rewritten
\begin{align}
 \label{acc-sf}
 {\mathcal F}^\mu_{\text{sf}}(\tau)
&=-\frac{\mu_0 q^2c^2\, a^\mu(\tau)}{4\pi a (a_1^2-a_2^2)} \,\bigg[I_1\bigg(\frac{c^2}{a a_1}\bigg) K_1\bigg(\frac{c^2}{a a_1}\bigg) - I_1\bigg(\frac{c^2}{a a_2}\bigg) K_1\bigg(\frac{c^2}{a a_2}\bigg)\bigg]\,.
\end{align}
In the limit to Bopp-Podolsky electrodynamics, $a_2\rightarrow 0$ and $a_1\rightarrow\ell$,
the self-force~\eqref{acc-sf} confirms the result found in \citep{Zayats,Perlick2015}.
Moreover, the self-force~\eqref{acc-sf} can be written in the form
\begin{align}
 \label{acc-sf-2}
 {\mathcal F}^\mu_{\text{sf}}(\tau)=-m_{\text{em}}(a) \, a^\mu(\tau)\,
\end{align}
with the electromagnetic mass of the moving particle
\begin{align}
 \label{m-em}
 m_{\text{em}}(a)
=\frac{\mu_0 q^2c^2}{4\pi a (a_1^2-a_2^2)} \, \bigg[I_1\bigg(\frac{c^2}{a a_1}\bigg) K_1\bigg(\frac{c^2}{a a_1}\bigg) - I_1\bigg(\frac{c^2}{a a_2}\bigg) K_1\bigg(\frac{c^2}{a a_2}\bigg)\bigg]\,,
\end{align}
which is the mass due to the interaction of the particle with its own electromagnetic field. 
It can be seen in Eq.~\eqref{acc-sf-2} that the self-force is directed against the acceleration of the charged particle.
The equation of motion, Eq.~\eqref{EOM}, can be written with Eq.~\eqref{acc-sf-2} as
\begin{align}
\label{EOM-2}
\big[m+m_{\text{em}}(a)\big]a^\mu(\tau)
=  {\mathcal F}^\mu_{\text{ext}}(\tau)\,
\end{align}
and we can put
\begin{align}
\label{m-exp}
m_{\text{exp}}=m+m_{\text{em}}(a)\,,
\end{align}
which may be considered as the experimentally measured mass.
Note that Eq.~\eqref{m-exp} is of the form of the so-called ``mass renormalization'' in electrodynamics (see, e.g.,  \citep{Barut,Rohrlich}).
The electromagnetic mass $m_{\text{em}}(a)$ is finite and depends on the constant acceleration $a$, in addition to the gradient parameters $a_1$ and $a_2$.
It possesses the following limits:
\begin{itemize}
\item
For vanishing particle acceleration,
it gives the electromagnetic rest mass of the particle:
\begin{align}
\label{lim-m2}
\lim_{a\to 0} m_{\text{em}}(a)=m_{\text{em}}(0)=\frac{\mu_0 q^2}{8\pi(a_1+a_2)}
\end{align}
appearing in the electrostatic self-energy, $U_{\text{self}}=m_{\text{em}}(0) c^2$, calculated in the framework of second gradient electromagnetostatics \citep{LL20}.
\item
For high particle acceleration, it goes to zero:
\begin{align}
\label{lim-m1}
\lim_{a\to \infty} m_{\text{em}}(a)=0\,.
\end{align}
\end{itemize}

\begin{figure}[t]\unitlength1cm
\vspace*{0.1cm}
\centerline{
\epsfig{figure=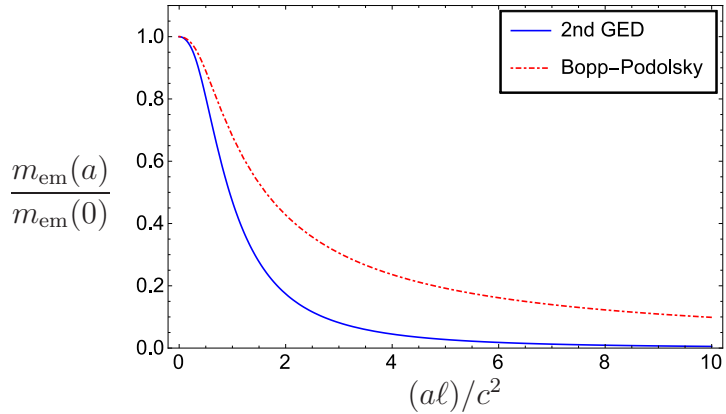,width=7.8cm}
\put(-4.2,-0.4){$(a \ell)/c^2 $}
\put(-9.5,2.3){$\displaystyle{\frac{m_{\text{em}}(a)}{m_{\text{em}}(0)}}$}
}
\caption{Plot of the electromagnetic mass of a uniformly accelerated charge $m_{\text{em}}(a)$
normalized to the electromagnetic rest mass $m_{\text{em}}(0)$ against $(a \ell)/c^2$
for $\ell = a_1 = 2 a_2$.}
\label{fig:m}
\end{figure}

The electromagnetic mass of a uniformly accelerated charge is plotted in Fig.~\ref{fig:m}.
The electromagnetic mass $m_{\text{em}}(a)$
represents the part of the mass of a uniformly accelerated particle which is due to the interaction with the electromagnetic field produced by this particle.

\section{Dispersion relations}
\label{Dispersion}

In this section, we analyze the wave propagation in the vacuum of second gradient electrodynamics.
We  show that the gradient terms affect the free space wave propagation, the structure of the vacuum
and the dispersion relations in second gradient electrodynamics.
In general, dispersion properties of spacetime are associated with noninstantaneous, nonlocal and gradient effects between the electromagnetic fields.

\begin{figure}[t]\unitlength1cm
\vspace*{0.1cm}
\centerline{
\epsfig{figure=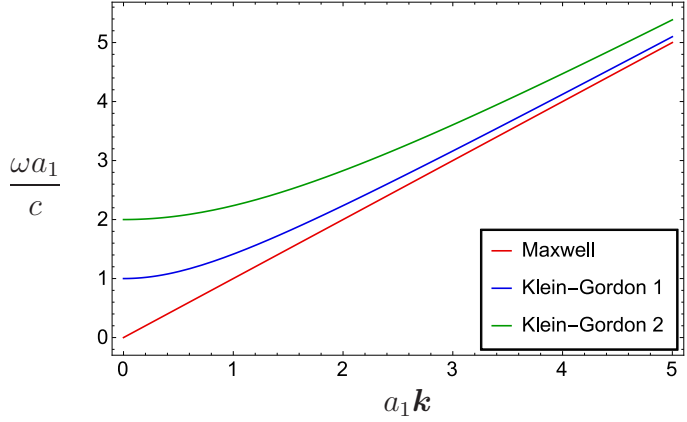,width=7.8cm}
\put(-4.0,-0.4){$a_1 \bm k $}
\put(-9,2.5){$\displaystyle{\frac{\omega a_1}{c}}$}
}
\caption{Dispersion relations in second gradient electrodynamics for the three modes:
Maxwell, Klein-Gordon 1 ($a_1$),
Klein-Gordon 2 ($a_2$) for $a_1=2a_2$.}
\label{fig:Dis}
\end{figure}

By means of the Fourier transform
\begin{align}
\label{A-FT}
A^\nu(x) =\int\limits_{\mathbb{R}^4} \tilde{A}^\nu(k)\, \exp({\ii k^\mu x_\mu})\, \d^4 k
\end{align}
and setting $j^\nu=0$ in the field equation~\eqref{EL-rel-A},
we find the dispersion relation
\begin{align}
\label{DR-rel}
-k^2
\big(1-a_1^2 k^2\big)
\big(1-a_2^2 k^2\big)=0\,.
\end{align}
Using
$k^\mu=(\omega/c, \bm k)$
and $k_\mu=(\omega/c, -\bm k)$, Eq.~\eqref{DR-rel} may be rewritten as
\begin{align}
\label{DR}
\bigg(\bm k^2-\frac{\omega^2}{c^2}\bigg)
\bigg(1+a_1^2 \bigg(\bm k^2-\frac{\omega^2}{c^2}\bigg)\bigg)
\bigg(1+a_2^2 \bigg(\bm k^2-\frac{\omega^2}{c^2}\bigg)\bigg)=0\,,
\end{align}
from which we can derive three different dispersion relations
for the propagation of electromagnetic waves in second gradient electrodynamics.
The first dispersion relation describes the typical non-dispersive waves of
Maxwell electrodynamics (dispersion relation of Maxwell-mode, see Fig.~\ref{fig:Dis}):
\begin{align}
\label{DR-M}
\omega^2=c^2 \bm k^2 \,.
\end{align}
The other two dispersion relations describe dispersive waves
of the Klein-Gordon equation
(two dispersion relations of Klein-Gordon-mode, see Fig.~\ref{fig:Dis}):
\begin{align}
\label{DR-KG1}
\omega^2&=c^2\bm k^2 +\frac{c^2}{a_1^2}\,,\\
\label{DR-KG2}
\omega^2&=c^2\bm k^2 +\frac{c^2}{a_2^2}\,.
\end{align}
On the one hand, the dispersion relations~\eqref{DR-KG1} and \eqref{DR-KG2} are known from the Klein-Gordon equation in quantum field theory describing
massive mesons.
On the other hand, the dispersion relations~\eqref{DR-KG1} and \eqref{DR-KG2} are known from waves in a collisionless plasma (e.g. \citep{Jackson,CD,Swanson}).
For each there is a minimum frequency $\omega_{a_1}=c/a_1$ or $\omega_{a_2}=c/a_2$, called the cutoff frequency (or plasma frequency),
below which the wave cannot propagate. Such behaviour is characteristic for waves propagating in  a plasma. 
For $\omega>\omega_{a_1}$ and $\omega>\omega_{a_2}$ the dispersive waves propagate without attenuation.

For the dispersive waves~\eqref{DR-KG1} and \eqref{DR-KG2}, the phase velocity $v_p=\omega/k$ and the group velocity $v_g=\pd \omega/\pd k$ are obtained as
\begin{align}
\label{V-KG1}
v_p&=c\,\sqrt{1+\frac{1}{a_1^2\bm k^2}}>c\,,\qquad
v_g=\frac{c}{\sqrt{1+\frac{1}{a_1^2\bm k^2}}}<c\,,
\\
\label{V-DR-KG2}
v_p&=c\,\sqrt{1+\frac{1}{a_2^2\bm k^2}}>c\,,\qquad
v_g=\frac{c}{\sqrt{1+\frac{1}{a_2^2\bm k^2}}}<c\,
\end{align}
satisfying $v_p v_g=c^2$. Since $v_p>c>v_g$, the dispersive waves demonstrate  normal dispersion at all frequencies above the corresponding
cutoff frequency.

Thus, we have found three kinds of waves in the vacuum of second gradient electrodynamics:
one mode of a non-dispersive wave like in Maxwell electrodynamics, and two
modes of dispersive waves like in a collisionless plasma.

It is noted that \citet{Santos2011} found two kinds of waves in the Bopp-Podolsky electrodynamics:
one mode of a non-dispersive wave as in Maxwell electrodynamics, and one
mode of a dispersive wave like in a collisionless plasma.
Therefore, second gradient electrodynamics possesses one mode of a dispersive wave more than Bopp-Podolsky electrodynamics,
due to the additional length scale parameter $a_2$.

\section{Conclusion}

\label{concl}

In this paper, we have developed the covariant theory of second gradient electrodynamics.
In this framework, we have derived the field equations, the energy-momentum tensor, the Lorentz force density, the retarded Green function,
the generalized Li\'enard-Wiechert potential, the electromagnetic field strength tensor of generalized  Li\'enard-Wiechert
type, the self-force as well as the equation of motion for a point charge, which is an integro-differential equation.
Because the Green function and its first-order derivative are regular,
covariant second gradient electrodynamics delivers a non-singular relativistic field theory of electromagnetism
including a covariant regularization based on higher-order partial differential equations.
Importantly, the electromagnetic field on the world line is not only finite, but also continuous, and thus allows for computation of the self-force without making additional assumptions for averaging.
Dispersion relations in second gradient electrodynamics were derived and one mode of non-dispersive wave and two modes of
dispersive waves were found.

Second gradient electrodynamics fits into the framework of axiomatic electrodynamics (see, e.g., \citep{Hehl})
with generalized constitutive relation (or electrodynamic spacetime relation).
The structure of covariant second gradient electrodynamics is given in Table~\ref{table}.
The basic structure of electrodynamics (e.g. homogeneous and inhomogeneous Maxwell equations)
keeps unchanged and only the constitutive relation (Maxwell-Lorentz spacetime relation) is generalized  in gradient electrodynamics.
The constitutive relation (or electrodynamic spacetime relation)
becomes a partial differential equation of fourth order in second gradient electrodynamics (see Table~\ref{table}).
Because the electrodynamic spacetime relation involves
time and space derivatives in gradient electrodynamics,  spacetime is temporally and spatially dispersive.

\begin{table}[t]
\caption{Axiomatic structure of second gradient electrodynamics.}
\begin{center}
\leavevmode
\begin{tabular}{ll}\hline\hline
Expression for & Second Gradient Electrodynamics \\
\hline
Continuity equation & $\pd_\nu j^\nu=0$\\
Field equation  & $\pd_\mu {\cal H}^{\mu\nu}= j^\nu$ \\
Bianchi identity & $ \pd_{[\alpha} F_{\mu\nu]}=0$ \\
Vacuum spacetime relation&
${\cal H}^{\mu\nu}
={\frac{1}{\mu_0}} \, \big(1+\ell_1^2\square+\ell_2^4\square^2\big)\,
F^{\mu\nu}$ \\
\hline\hline
\end{tabular}
\end{center}
\label{table}
\end{table}

\section*{Acknowledgement}
M.L. gratefully acknowledges the grant from the Deutsche Forschungsgemeinschaft (Grant No. La1974/4-1).

\begin{appendix}
\section{Relation to second gradient electrodynamics in vector form}
\label{appendixA}
\setcounter{equation}{0}
\renewcommand{\theequation}{\thesection.\arabic{equation}}

In this appendix, we derive the vector form from the covariant form of second gradient electrodynamics.
The electromagnetic field strength tensor, which can be given in terms of the electric field strength vector $\bm E$ and the magnetic
field strength vector $\bm B$, reads  in $4\times 4$-matrix representation (see, e.g., \citep{Freeman})
\begin{align}
\label{F-M}
F_{\mu\nu}=\begin{pmatrix}
0 & E_x/c &  E_y/c& E_z/c\\
-E_x/c & 0&  -B_z& B_y\\
-E_y/c &B_z& 0 & -B_x\\
-E_z/c & -B_y& B_x &0
\end{pmatrix}\,,
\quad
F^{\mu\nu}=\begin{pmatrix}
0 & -E_x/c &  -E_y/c& -E_z/c\\
E_x/c & 0&  -B_z& B_y\\
E_y/c &B_z& 0 & -B_x\\
E_z/c & -B_y& B_x &0
\end{pmatrix}\,.
\end{align}
Using
\begin{align}
\label{FF}
-\frac{1}{4\mu_0}\, F^{\mu\nu}F_{\mu\nu}=\frac{1}{2}\,\Big(\varepsilon_0\, \bm E \cdot \bm E -\frac{1}{\mu_0}\, \bm B \cdot \bm B\Big)
\end{align}
and the contravariant current density four-vector $j^\nu=(c\rho,\bm j)$,
the covariant current density four-vector $j_\nu=(c\rho,-\bm j)$,
the covariant electromagnetic potential four-vector $A_\mu=(\phi/c,-\bm A)$,
the contravariant electromagnetic potential four-vector $A^\mu=(\phi/c,\bm A)$,
$\pd_\mu=(\pd_t /c,\nabla)$ and  $\pd^\mu=(\pd_t /c,-\nabla)$,
the Lagrangian density~\eqref{L-BP-rel} can also be expressed in vector form as given in~\citep{Lazar20}
\begin{align}
\label{L-BP}
{\cal L_{\text{grad}}}&=
\frac{\varepsilon_0}{2}
\Big(\bm E \cdot \bm E
+\ell_1^2
\Big[
\nabla \bm E :\nabla \bm E
-\frac{1}{c^2}\, \pd_t \bm E \cdot \pd_t \bm E
\Big]\nonumber\\
&\quad
+\ell_2^4
\Big[
\nabla\nabla \bm E \mathbin{\vdots} \nabla\nabla \bm E
-\frac{2}{c^2}\, \pd_{t} \nabla \bm E : \pd_{t} \nabla \bm E
+\frac{1}{c^4}\, \pd_{tt} \bm E \cdot \pd_{tt} \bm E
\Big]
\Big)
\nonumber\\
&\quad
-\frac{1}{2\mu_0 }
\Big(\bm B \cdot \bm B
+\ell_1^2
\Big[
\nabla \bm B :\nabla \bm B
-\frac{1}{c^2}\, \pd_t \bm B \cdot \pd_t \bm B
\Big]\nonumber\\
&\quad
+\ell_2^4
\Big[
\nabla\nabla \bm B \mathbin{\vdots} \nabla\nabla \bm B
-\frac{2}{c^2}\, \pd_{t} \nabla \bm B : \pd_{t} \nabla \bm B
+\frac{1}{c^4}\, \pd_{tt} \bm B \cdot \pd_{tt} \bm B
\Big]
\Big)
\nonumber\\
&\quad
-\rho\phi+\bm j\cdot \bm A
\end{align}
with the notation:
$ \bm E \cdot\bm E =E_i E_i$,
$ \nabla \bm E :\nabla \bm E =\pd_j E_i \pd_j E_i$ and
$ \nabla\nabla \bm E \mathbin{\vdots} \nabla\nabla \bm E
=\pd_k\pd_j E_i \pd_k \pd_j E_i$. 
Here, $\nabla$ denotes the del operator,
$\pd_t$ is the differentiation with respect to time $t$,
$\rho$ is the electric charge density,
$\bm j$ is the electric current density vector,
$\phi$ is the electric scalar potential
and $\bm A$ is the  electromagnetic vector potential.
In the vector form, Eq.~\eqref{F} gives
\begin{align}
\label{E}
\BE&=-\nabla \phi-\pd_t \bm A\,,\\
\label{B}
\BB&=\nabla\times \bm A\,,
\end{align}
while the Bianchi identity~\eqref{BI-rel} gives rise to the homogeneous Maxwell equations for $(\bm E, \bm B)$,
\begin{align}
\label{BI-1}
\nabla\times\BE+\pd_t\BB&=0
\,,\\
\label{BI-2}
\nabla\cdot \BB&=0\,,
\end{align}
and the inhomogeneous Maxwell equation~\eqref{EL-rel-F}
reduces to (see also~\citep{Lazar20})
\begin{align}
\label{EL-1}
 L(\square)\,
\nabla\cdot \bm E&=\frac{1}{\varepsilon_0}\,\rho\,,\\
\label{EL-2}
 L(\square)
\Big(
\nabla\times\bm B-\frac{1}{c^2}\,\pd_t\bm E\Big)&=\mu_0\,\BJ\,.
\end{align}

\end{appendix}

\end{document}